\documentclass[aps,prl,preprint,groupedaddress]{revtex4-1}
\usepackage{amsmath,amssymb}
\usepackage[dvips]{graphicx}

\begin{document}

\title{Neutrino diffraction induced by many-body interaction}
\author{Kenzo Ishikawa and Yutaka Tobita}
\affiliation{Department of Physics, Faculty of Science, Hokkaido
University, Sapporo 060-0810, Japan}

\date{\today}
\begin{abstract} 
The neutrino produced in the pion decay reveals a new diffraction 
 phenomenon  due to  many-body interactions in  an intermediate 
time  region when wave functions of the
 parent and daughters overlap.
Because of  diffraction,    the probability to detect the neutrino
 involves  a large finite-size correction  that depends 
on  the neutrino mass, $m_{\nu}$ and energy, $E_{\nu}$, the speed of light,
 $c$, and the distance L between the positions of the initial pion 
and  final neutrino,  ${m_{\nu}^2c^4 \text{L}}/{ (2E_{\nu}\hbar)}$. 
 The correction vanishes for  the charged leptons and is finite for  the
 neutrino   at a macroscopic distance, L, 
of near-detector regions in  ground  experiments. A new
 method  for   determining the absolute neutrino mass is proposed.

\end{abstract}
\maketitle
{\bf 1 Neutrino interference.} Interference  of photons,
electrons, neutrons, and other heavy
elements are important for confirming   quantum mechanics and other basic
principles. A wave composed of many components of different kinetic
energies  behaves  non-uniformly  in space.
In the above cases, they are formed by   potential  energy.  We present   
a diffraction  theory due to  many-body 
interactions that provides  a varying  kinetic energy to  the 
many-body system in a finite time.  
A neutrino produced in  pion decay reveals the  wave nature, and the 
 probability for  detecting the neutrino displays  
 diffraction.  Consequently the absolute value of the currently unkown 
neutrino mass could be deduced from  the
unique interference pattern of this diffraction phenomenon.

 A neutrino interacts extremely weakly with matter. Being undisturbed by
 matter, the neutrino behaves purely as a  quantum mechanical wave with  a
negligible one-particle potential, except in  case of degenerate flavor
states \cite{MSW-1,MSW-2}.  Instead of  potential energy,    
 interaction energy carried by a weak Hamiltonian  becomes finite 
when   the wave functions of the pion and  decay product
overlap. Then,   kinetic energy  becomes different from that of the initial
state and   varies with time, because  total energy is constant. 
This 
system  shows   a non-uniform spatial behavior,  called neutrino 
diffraction, similar to the above cases of ordinarly particles.   Having 
its origin in the  weak
 Hamiltonian,  neutrino diffraction appears   in vacuum and has  
universal properties.  The diffraction pattern  is easily observed 
without an obstacle or potential  in the time region $\text T \leq
\tau$, where T is
the  time interval between  initial and final states, and $\tau$ is 
the  mean life-time for  a large number of events.

Consider the system described by the  Hamiltonian $H=H_0+H_1$, where
$H_0$ is a
bi-linear form of fields, and $H_1$ is a higher-order polynomial that causes
many-body interactions.  Kinetic energy is defined by eigenvalue 
$E$ of $H_0$. 
The  Schr\"{o}dinger  equation, $i{\partial \over \partial
t}|\psi(t)\rangle=(H_0+H_1)|\psi(t)\rangle$ is solved using $H_1$
 and an initial state $|\psi^{(0)}\rangle$ of 
the  kinetic energy, $E_0$, in the 
interaction picture by $|\tilde \psi(t)\rangle=\mathcal{T}\int_0^t dt'
e^{-i\tilde H_1(t')}|\tilde \psi^{(0)}\rangle$, where $\mathcal{T}$
denotes   the
time-ordered product.
Hence,  the wave
function  $|\tilde \psi(\infty) \rangle $ 
is written 
in
the following form; 
\begin{eqnarray}
|\tilde \psi(\infty) \rangle=a(\infty)|\tilde\psi^{(0)}\rangle+2\pi \int d{\beta} \delta(\omega)|\tilde\beta \rangle \langle \tilde\beta
 |\tilde S |\tilde\psi^{(0)} \rangle
\end{eqnarray} 
with a reduced matrix $\tilde S $, where $H_0|\tilde\beta
 \rangle=E_{\beta}|\tilde\beta \rangle,\ \omega=E_{\beta}-E_0$, and 
 $a(\infty)$ is a constant.
 The state $|\tilde\beta \rangle$  has the kinetic energy of the 
initial state,  $E_0$. Accordingly, this
state has the  property  of free
particles. 
Kinetic energy   is conserved
in  the asymptotic regions $t\rightarrow \pm\infty$, and 
a scattering matrix $S[\infty]$ satisfies 
$[S[\infty],H_0]=0$. 
Now, at  finite $t$, the wave
function is written as follows:
\begin{eqnarray}
|\tilde\psi(t) \rangle=a(t)|\tilde\psi^{(0)}(t)\rangle+\int d {\beta}{e^{i\omega t}-1 \over \omega}|\tilde\beta \rangle \langle \tilde\beta
 |\tilde S |\tilde\psi^{(0)} \rangle,
\label{finite-tamplitude}
\end{eqnarray}
and  it is a superposition of  the states of the kinetic energy, $E_0$ and  
 $E_{\beta}\geq 0$, with a time dependent weight. 
The interaction energy, $\langle \tilde
\psi(t)|\tilde H_1|\tilde \psi(t)\rangle  $, does not vanish  in the 
region of  finite   
$\langle \tilde \psi^{(0)}(t)|\tilde H_1|\tilde \beta \rangle $. 
Total energy satisfies  the following conditon: 
$\langle \tilde
\psi(t)|\tilde H|\tilde \psi(t)\rangle=E_0$.

At  finite t, $|\tilde \psi(t) \rangle$ has a varying  kinetic energy
that  free particles do not possess.  Consequently,  $|\tilde \psi(t)
\rangle$ retains 
its wave nature, and  the   
probability to detect a particle in the final state
becomes  dependent on    a  time interval, which we call  finite-size 
correction.
To observe this correction, an S-matrix  $S[\text T]$  defined according to the 
boundary condition at the  time interval T is used.  Because  wave packets 
  localize around center positions and     satisfy the asymptotic boundary
conditions \cite {LSZ,Low} of scattering experiments,   $S[\text T]$ is
defined using wave packets.   
 $S[\text T]$ is constructed   with   M{\o}ller operators at  finite T,
 $\Omega_{\pm}(\text{T})$, as $S[\text T]=\Omega_{-}^{\dagger}(\text T)\Omega_{+}(\text
T)$.  The term $\Omega_{\pm}(\text{T})$  is expressed  in the form   $\Omega_{\pm}(\text T)=\lim_{t
 \rightarrow \mp \text T/2}e^{iHt}e^{-iH_0t}$.  From this expression,     
$S[\text T]$ satisfies the following equation:
\begin{align}
&[S[\text T],H_0]=  i\left\{\frac{\partial}{\partial
	\text{T}}\Omega^{\dagger}_{-}(\text{T})\right\}\Omega_{+}(\text{T})-
	i\Omega^{\dagger}_{-}(\text{T}){\partial \over \partial \text{T}}
	\Omega_{+}(\text{T}).
\label{commutation-relation}
\end{align}
Thus,   kinetic energy is not conserved at  finite T.
A  matrix
element of $S[\text T]$ between eigenstates $|\alpha \rangle$ 
and  $|\beta
\rangle $ of eigenvalues $E_{\alpha}$ and $E_{\beta}$ respectively, 
$ \langle \beta|S[\text T]|\alpha \rangle$, has  the  components
of $E_{\beta}=E_{\alpha}$ and $E_{\beta}\neq E_{\alpha}$.
 At $\text T \rightarrow \infty$, only the former terms remain,  and 
the latter terms  at  finite T  give  finite-size correction.  

A neutrino wave packet  \cite {Ishikawa-Shimomura,Ishikawa-Tobita-ptp,Ishikawa-Tobita}
expresses   a nucleon wave function in a nucleus with which the neutrino
interacts  and   is  
well-localized \cite{Kayser,Giunti,Nussinov,Kiers,Stodolsky,Lipkin,Akhmedov,Asahara}.   
 The mass-squared differences, $\delta m^2_\nu$, are extremely small   \cite
 {particle-data,Tritium,WMAP-neutrino}, thus, we study
 a situation in which the  mass-squared average, ${\bar m}^2_\nu$,
 satisfies, ${\bar m}^2_\nu \gg \delta
 m^2_\nu $, and presents the  one flavor case first. Extensions to general cases are 
straightforward. 

{\bf 2 Position-dependent probability.}  

 Now, we find the finite-size correction of the probability to detect the
 neutrino in the pion
 decay with $S[\text T]$.  $H_0$ is the free Hamiltonian of the pion, charged lepton, and 
 neutrino; and  
$H_1=g\int d{\vec x} \partial_{\mu} \varphi (V-A)_{lepton}^{\mu}$, where $\varphi(x)$, 
$V^{\mu}(x)$, and $A^{\mu}(x)$ are pion field, lepton's vector,  and 
axial-vector currents respectively. 
The term  $|\psi^{(0)}\rangle $ is a one-pion state, and $|\beta
 \rangle$ is a two- particle state composed of a charged lepton and neutrino. 
 For   a  pion of momentum $p_{\pi}$ prepared  at  $\text{T}_{\pi}$,
 the amplitude for a
 neutrino of $p_{\nu}$  to be 
detected at $(\text{T}_{\nu},\vec{\text{X}}_{\nu})$ and
 a muon of $p_{\mu}$ to be un-detected  in the lowest order
 of $H_1$,  $T=\int d^4x \, \langle {\mu},{\nu}
 |H_{1}(x)| \pi \rangle$, 
is written in terms of   Dirac spinors  as follows: 
\begin{align}
T = \int d^4xd{\vec k}_{\nu}
\,N\langle 0 |\varphi_\pi(x)|\pi \rangle 
\bar{u}({\vec p}_{\mu}) (1 - \gamma_5)\nu({\vec k}_{\nu}) \nonumber\\
\times e^{ip_{\mu}\cdot x + 
ik_\nu\cdot(x - \text{X}_\nu)
 -\frac{\sigma_{\nu}}{2}({\vec k}_{\nu}-{\vec p}_{\nu})^2},
\end{align}
where  
$N=igm_{\mu}\left({\sigma_\nu/\pi}\right)^{\frac{4}{3}}\left({m_{\mu}m_{\nu}}/{
 E_{\mu}E_{\nu}}\right)^{\frac{1}{2}}$, and  the four-dimensional
 coordinate, $x$, has the components $(t,{\vec x})$, and  $t$ is
 integrated over the region $\text{T}_{\pi} \leq t$. 
In this study, a Gaussian form is assumed for  simplicity.  Finite-size  correction  has a universal property 
that is common to general wave packets. The size of the
wave packet, ${\sigma_{\nu}}$,  is estimated later. 
The amplitude $T$  satisfies the boundary condition at  finite 
$\text{T} = \text{T}_\nu - \text{T}_\pi$.

By integrating  ${\vec k}_{\nu}$, we obtain  the Gaussian function of
${\vec x}-{\vec x}_0$, which vanishes
at large $|{\vec x}-{\vec x}_0|$, where ${\vec x}_0$ is the center
coordinate to be  expressed later, and satisfies  the asymptotic boundary
condition. We express  an integration of $|T|^2$  over ${\vec p}_{\mu}$ with
 a correlation function of coordinates.
 After  spin summations, we have the following expressions: 
 \begin{align}
&P=\int
\frac{d {\vec p}_{\mu}}{(2\pi)^3}\sum_{\text {spin}}|T|^2 = \frac{C}{E_\nu}\int d^4x_1 d^4x_2 
e^{-\frac{1}{2\sigma_\nu}\sum_i ({\vec x}_i-\vec{x}_i^{\,0})^2}\nonumber\\
&\hspace{120pt}\times \Delta_{\pi,\mu}(\delta x)e^{i \phi(\delta x)}e^{-{t_1+t_2 \over \tau}},
\label{probability-correlation} 
\\
&\Delta_{\pi,\mu} (\delta x)=
 {\frac{1}{(2\pi)^3}}\int
{d {\vec p}_{\mu} \over E({\vec p}_{\mu})}  (p_{\mu}\!\cdot\! p_{\nu})
 e^{-i(p_{\pi}-p_{\mu})\cdot\delta x }, 
\label{pi-mucorrelation}
\end{align}
where  $\tau$ is pion's  life-time,  $C=g^2m_{\mu}^2
\left({4\pi}/{\sigma_{\nu}}\right)^{\frac{3}{2}}V^{-1}$, $V$ is
a  normalization volume for the initial pion, $\vec{x}_i^{\,0} = \vec{\text{X}}_{\nu} + {\vec
v}_\nu(t_i-\text{T}_{\nu})$, $\delta x
=x_1-x_2$, and  $\phi(\delta x)=p_{\nu}\!\cdot\!\delta x $.
In Eq.\,$(\ref{pi-mucorrelation})$,  muon momentum is  integrated 
in the entire  positive energy region so that  
Eq.\,$(\ref{probability-correlation}  )$ can  agree with the original probability.

{\bf 3 Light-cone singularity.}

By using the  new variable  $q=p_{\mu}-p_{\pi}$ that is a conjugate to $\delta
 x$, we write   $\Delta_{\pi,\mu}(\delta x)$ as the  sum of    
integrals  over  regions $0 \leq q^0$, and   $-p_{\pi}^0 \leq
q^0 \leq 0$. Kinetic energy is conserved in the latter integral  and
 not conserved in the former. Thus, they  correspond to the asymptotic
 value and  finite-size correction.   
The former integral is expressed as, $ \left[p_{\pi}\! \cdot\!
 p_{\nu} -ip_{\nu}\!\cdot\! ({\partial \over
 \partial \delta x} )\right] \tilde I_1$, and  the four-dimensional
 integral is given as follows:  
\begin{align}
\tilde I_1=\int d^4 q \,  \frac{\theta(q^0)}{4\pi^4}\text {Im}\left[1 \over
 q^2+2p_{\pi}\!\cdot\! q+{\tilde m}^2-i\epsilon\right] e^{iq \cdot \delta x
 }\nonumber,
\end{align}
and  ${\tilde m}^2=m_{\pi}^2-m_{\mu}^2$. 
By expanding the denominator with $p_{\pi}\!\cdot\! q$, we have 
an  expression using  the light-cone singularity \cite{Wilson-OPE}, $\delta({\delta
x}^2)$,  and the less singular and normal  terms  that  are described with 
Bessel functions.  
The latter integral, $I_2$,  
 has no singularity. 
By adding both terms, we have the following expressions: 
\begin{align}
&\Delta_{\pi,\mu}(\delta x)=2i
\left\{p_{\pi}\! \cdot\! p_{\nu}-ip_{\nu}\!\cdot\!\left(\frac{\partial}{
 \partial \delta x}\right)\right\}\nonumber\\
&\times  \left[ D_{\tilde m}\left(-i\frac{\partial}{\partial \delta x}\right) 
\left( \frac{\epsilon(\delta t)}{4\pi}\delta(\lambda)  
+f_{short}\right) +I_2\right],\label{muon-correlation-total}
\end{align}
where $\lambda=(\delta x)^2$, $D_{\tilde m}(-i\frac{\partial}{\partial
\delta x})$
$=$
$\sum_l$
$ (1 /l!)\bigl(2p_{\pi}\!\cdot\!(-i{\partial \over \partial \delta x})$
$\frac{\partial}{\partial {\tilde m}^2}\bigr)^l$, 
and $f_{short}$ is expressed with Bessel functions \cite{Wilson-OPE,Ishikawa-tobita-lightcone}.

{\bf Integration of  coordinates.} Next,
Eq.\,$(\ref{muon-correlation-total})$ is substituted into
Eq.\,$(\ref{probability-correlation})$, and  ${\vec x}_1$ and ${\vec x}_2$ are integrated. 
The light-cone singularity, $\frac{\epsilon(\delta
t)}{4\pi}\delta(\lambda)$, leads  to the slowly oscillating  term, $J_{\delta(\lambda)}$:       
\begin{align}
J_{\delta(\lambda)}=C_{\delta(\lambda)}
\frac{\epsilon(\delta t)}{|\delta t|
 }e^{i\bar \phi_c(\delta t)},\label{lightcone-integration2-2}
\end{align}
where $C_{\delta(\lambda)}={(\sigma_{\nu}\pi)}^{\frac{3}{2}}
 {\sigma_{\nu}}/{2}$, and $\bar\phi_c(\delta t)=\omega_{\nu} \delta t=\delta t~{m_{\nu}^2c^4 /(
 2E_{\nu})}$. The phase $\phi(\delta x)$ of
 Eq.\,$(\ref{probability-correlation})$ becomes  the small phase 
$\bar \phi_c(\delta t)$ of Eq.\,$(\ref{lightcone-integration2-2})$ at the
 light cone $\lambda=0$. 
The next singular 
term  
becomes  much smaller than that in the present parameter region, 
and the  normal  terms 
  oscillate  or decrease  rapidly with $\lambda$ and those of
 ${\vec r}
\approx 0$  contribute to the oscillation or decreas.  Hence, the  spreading effect is
negligible. 
The terms $f_{short } $ and $I_2$ in
Eq.\,$(\ref{muon-correlation-total})$  lead to rapidly  oscillating or
decreasing terms which we denote by $\tilde L$.


Finally,  we integrate $t_1$ and $t_2$ over the finite region, 
$0\leq t_i \leq
\text T$: 
\begin{align}
\label{total-probability2}
P=N_1\int_0^{\text T} dt_1 dt_2\biggl[    
\frac{\epsilon(\delta t)}{|\delta t|}e^{i {\bar \phi_c(\delta t)}}+
 \tilde L \biggr] e^{-{t_1+t_2 \over \tau}},
\end{align}
where $N_1=ig^2 m_{\mu}^2
\pi^3{\sigma_{\nu}}(8p_{\pi}\!
 \cdot\! p_{\nu}/E_{\nu})V^{-1}$.
In  most  of the places, the  neutrino mass 
 is neglected compared to $\tilde{m}^2$, $p_{\pi}\!\cdot\! p_{\nu}$ and
$\sigma_{\nu}^{-1}$,  except for the slow phase $\bar \phi_c(\delta t)$. The 
first term in
Eq.\,$(\ref{total-probability2})$ oscillates slowly with  $\delta t$, 
 and the remaining terms oscillate or decrease rapidly.  They  are
 clearly separated. 
\begin{eqnarray}
& &i  \int_0^{\text{T}} dt_1 dt_2  \frac{\epsilon(\delta t)}{|\delta t|}e^{i {\omega_{\nu}}\delta t }  e^{-{t_1+t_2 \over \tau}} 
=  (\tilde  g(\omega_{\nu}\text{T})-\tilde g_0)\ 
\label{probability1} 
\end{eqnarray}
The  first term slowly approaches $\tilde g_0$ with T, where     $\tilde
g(\omega_{\nu}\text{T})$ satisfies
${\partial \over \partial \text{T}}\tilde g(\omega_{\nu}\text{T})|_{\text{T}=0}
 = -\omega_{\nu}$, and  $\tilde g({\infty})=0$. 
The constant $\tilde g_0$   cancels  the short-range term, $\tilde L$, 
in   Eq.\,$(\ref{total-probability2})$.
Here $\tilde g(\omega_\nu\text{T}) $ is generated by
the light-cone singularity, 
and its effect remains within  a macroscopic 
distance of  order $\frac{2c\hbar E_{\nu}}{m_{\nu}^2c^4}$. We call 
this as the {\bf diffraction} term. 
From the  last  term of
Eq.\,$(\ref{total-probability2})$, $G_0(\text T)$ is defined as  $
i \int dt_1 dt_2 \tilde
 L(\delta t)e^{-{t_1+t_2 \over \tau}} = G_0(\text T)+\tilde g_0$.
 Because of rapid oscillation in $\delta t$,  the {\bf normal} term,
 $G_0(\text T)$, receives  
 contributions  from 
 the microscopic  $|\delta t|$ region,  is proportional to T in 
the region $\text T < \tau$, and
approaches a constant for  large T.

The present method of extracting the light-cone singularity is valid 
if  the series $ D_m(-{\partial \over \partial \delta x}) f_{short}$ 
converges.
 This  condition is  fulfilled,   \cite{Ishikawa-tobita-lightcone},     
in the region $2p_{\pi}\! \cdot\! p_{\nu}\leq {\tilde
m}^2 $, and
 the  series  rapidly oscillates.
Outside  this region, the method is not applicable, and
$\Delta_{\pi,\mu}(\delta x)$  has no light-cone singularity and  has 
only the short-range term.

{\bf 4 Total probability that depends on  time interval.}

From the integration of   the neutrino coordinate $\vec{\text{X}}_{\nu}$,
 the total volume emerges and   cancels  with 
$V^{-1}$. The total
probability becomes    
\begin{align}
\label{probability-3}
P=N_2\int \frac{d\vec{p}_{\nu}}{(2\pi)^3}
\frac{p_{\pi}\! \cdot\! p_{\nu}}{E_\nu}
 \left[\tilde g(\omega_{\nu}\text{T}) 
 +G_0(\text T) \right],
\end{align}
where $N_2 = 8g^2 m_{\mu}^2\sigma_\nu$, and $\text{L} =
c\text{T}$ is the length of the decay region. Because  $G_0(\text T)$ and $\tilde
g(\omega_{\nu} \text T)$  have  their origins in the
conserving and non-conserving terms  of
 kinetic energy, respectively,  $p_{\pi}\!\cdot\!
 p_{\nu}={\tilde{m}^2/2}$  in
$G_0(\text T)$ but not in $\tilde g(\omega_{\nu}\text T)$.
 \begin{figure}[t]%
  \begin{center}
   \includegraphics[angle=-90,scale=.35]{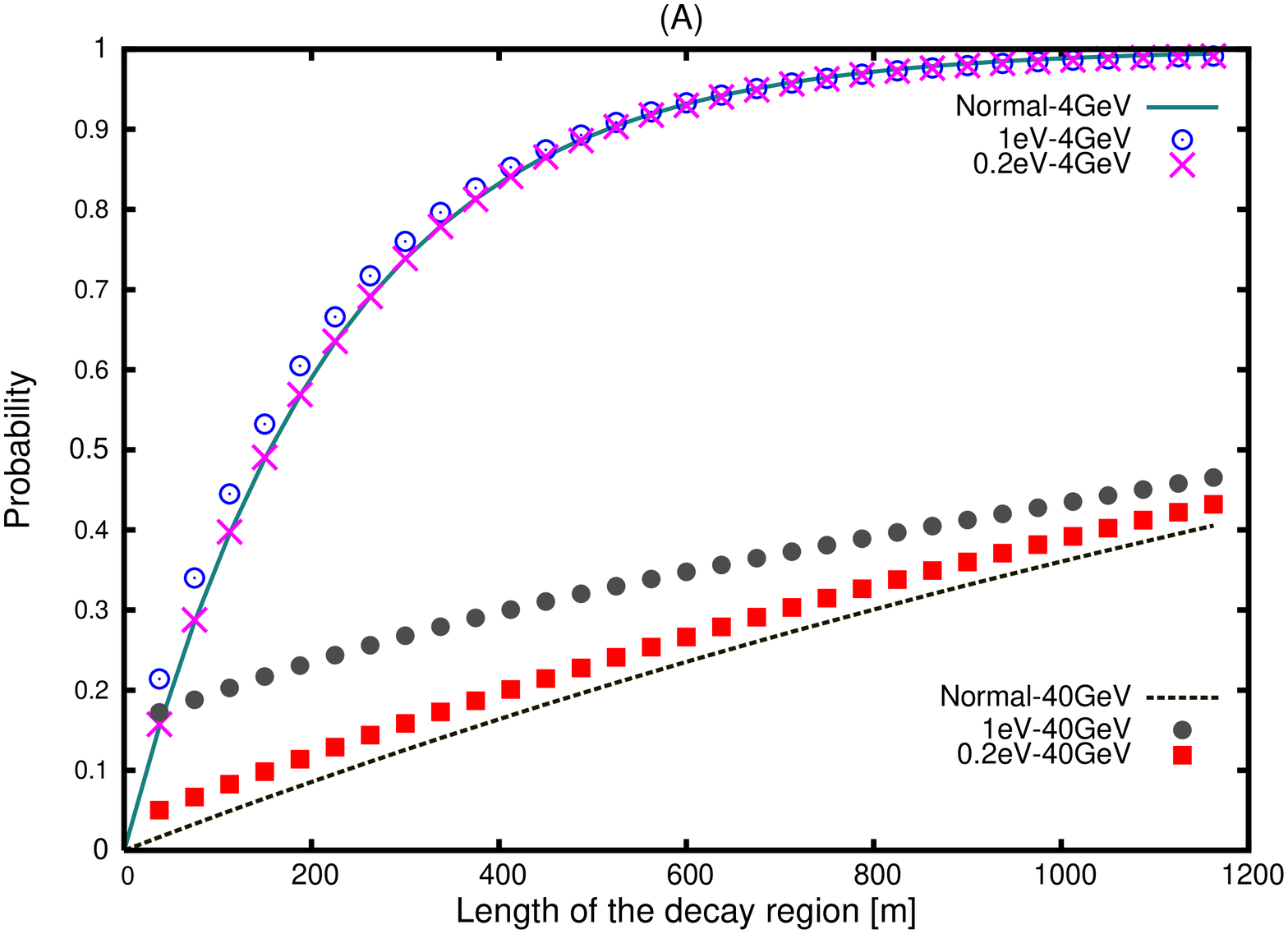}\\
   \includegraphics[angle=-90,scale=.35]{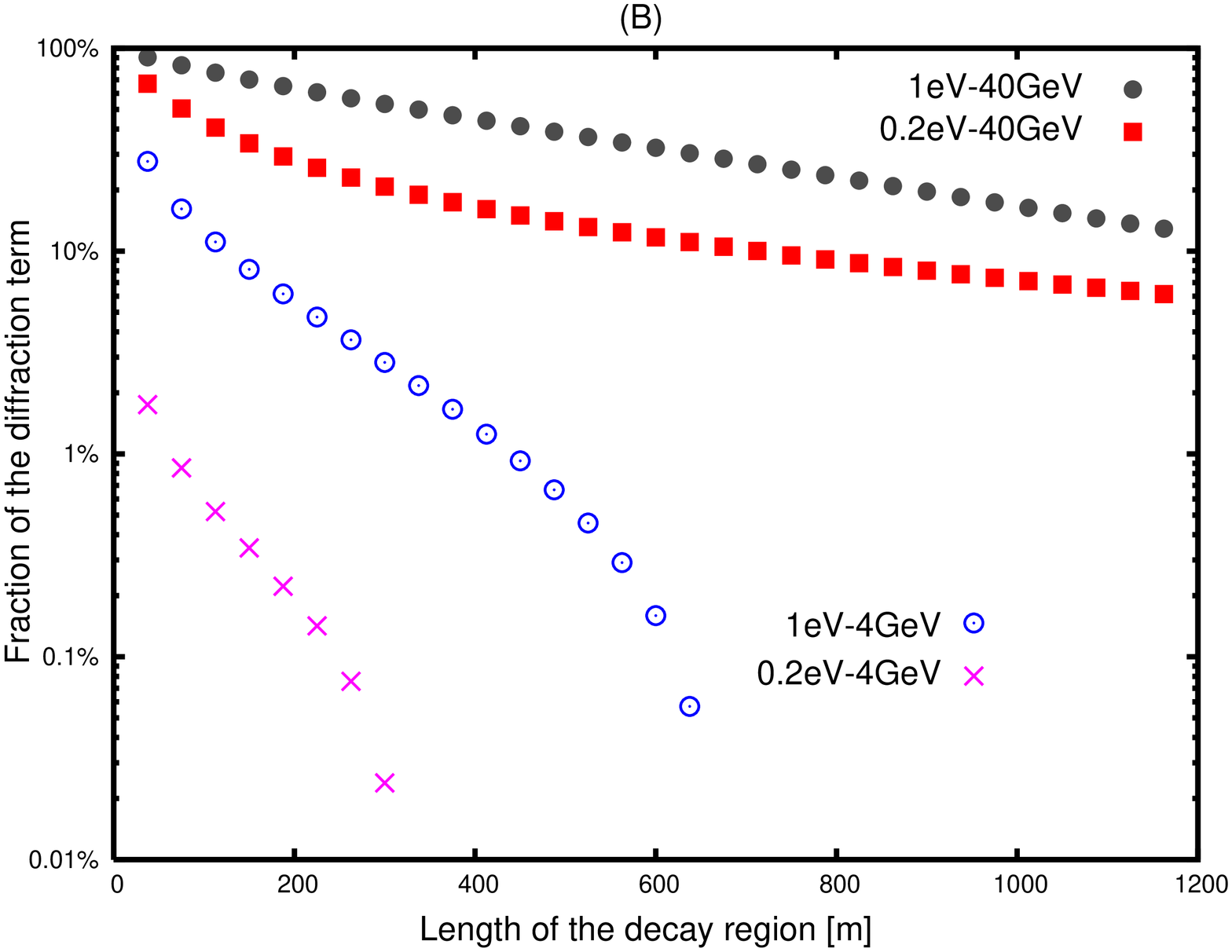}%
  \end{center}
  \caption{Detection probabilities of the neutrino  for $E_{\pi}=4$ and $40$ [GeV]
  at   distance L.   In (A),  solid  green ($4$ [GeV]) 
and  dotted black line  ($40$ [GeV]) represent normal and and
  diffraction 
terms,respectively, for  for 
$m_{\nu}$ of $0.2$ and $1.0$ [eV] is mentioned  on top of  normal
  terms. Values   are normalized 
to one  at $\text{L}=\infty$.
 In (B),   fractions of 
  diffraction terms that    vary with
   pion's energy and  neutrino mass are shown. The horizontal axis 
  represents  distance in~[m]. 
   Neutrino energy is 700~[MeV].}
  \label{fig:total-int-1}
 \end{figure}%
By integrating the neutrino angle, we find that  the normal 
  term is independent of  $\sigma_{\nu}$  \cite{Stodolsky} and agrees with
  the  value computed via   plane waves. 
However,   
 $\tilde g(\omega_{\nu}\text{T})$ 
is present in the kinematical region; i.e.,  
$|\vec{p}_{\nu}|(E_{\pi}-|\vec{p}_{\pi}|)\leq p_{\pi}\!\cdot\! p_{\nu}
\leq {\tilde{m}^2/2}$ from the convergence condition,
and $\tilde g(\omega_{\nu}\text T)$ is integrated in this region. 
This  is 
slightly different from $p_{\pi}\!\cdot p_{\nu}={\tilde{m}^2/2}$; hence,
  the latter  region cannot be distinguished from the former. Therefore,
we add the two  terms. Total probability thus obtained 
is
 presented  in 
Fig.\,\ref{fig:total-int-1} for 
 $m_{\nu}=1,\ 0.2\,[\text{eV}/c^2]$,  $E_\pi = 4,\ 40$\,[GeV],
  and $E_\nu = 700$ [MeV]. The size
 of the nucleus of a mass number, $A$, is used for the wave packet, $\sigma_{\nu}=
 A^{\frac{2}{3}}/m_{\pi}^2$, and    $\sigma_{\nu}= 6.4/m_{\pi}^2$ 
 is used for the  evaluation of the ${}^{16}$O 
nucleus.    From
  Fig.\,\ref{fig:total-int-1},  we can observe    that the diffraction term
  becomes finite in  ${\text{L}/c \leq \tau}$, where the wave functions of the
  initial and final states overlap. Their fractions  at $\text{L} \approx 0$ vary from  
$0.02$ for $m_{\nu}=0.2$ [eV] to $0.2$ for $m_{\nu}=1.0$ [eV]
  at $4$ [GeV], and  they become approximately    
$ 1.0$ for $m_{\nu}=0.2$ [eV] and $m_{\nu}=1.0$ [eV]   at $40$
  [GeV]. They decrease rapidly with L at $4$ [GeV] and slowly at  $40$
  [GeV], because the life-time is longer for  the latter energy. 
The diffraction term  slowly varies  with L in the  high energy region,
 in which   the life-time effect becomes negligible, and 
a typical length, $\text{L}_0$, for this  behavior  is given as  
$\text{L}_0~[\text{m}] ={2E_{\nu} \hbar c / (m_{\nu}^2c^4) }= 400\times {E_{\nu}[\text{GeV}]/
 m_{\nu}^2[\text{eV}^2/c^4]}$. In the experiments,
  neutrino's 
energy is measured with uncertainty $\Delta E_{\nu}$,
 which is of the 
order $0.1 \times E_{\nu}$ and  is $100$\,[MeV] for  energy
$1$\,[GeV]. Diffraction components  are
 almost constant in this energy range. 
For a larger energy uncertainty, the computation  is easily made  using
Eq.\,(\ref{probability-3}).   
Hence, the diffraction component  is observable  if  $m_{\nu} \geq
  0.2\,[\text{eV}/c^2]$ using the near detector, but it  is not 
 observable if  $m_{\nu} \leq 0.1\,[\text{eV}/c^2]$ using the muon neutrino. In
 the latter  case,  an
electron neutrino may be  used.  

The  process described using $S[\text{T}]$ has   the
 total probability  same as that shown in
 Eq.\,$(\ref{probability-3})$. In the same
experiment, the detection rate of the muon,  after the neutrinos are 
integrated,  
 has  the  same excess value. Ordinary experiments of observing the muon,
 however, do not consider  the neutrino and are described 
by a different 
$S[\text{T}']$, which satisfies the boundary condition for the muon, and 
 $\text{T}'=\text{T}_{\mu}-\text{T}_{\pi}$ is the 
time interval for  muon observation. 
The 
probability to detect a muon  is computed with a free neutrino, and then
 it is expressed in the form of Eq.\,$(\ref{probability-3})$ with
$\omega_{\nu} \rightarrow \omega_{\mu}= {m_{\mu}^2c^4/(2E_{\mu}\hbar)}$. 
Because  the muon is   heavy, $\omega_{\mu}\text{T}'$ becomes very
 large,  and $\tilde g(\omega_{\mu}\text{T}')$ vanishes at at 
 macroscopic $\text{T}$.  Thus,  the 
probability of detecting the muon  is not modified,  and  it agrees with 
the normal
term. 
The light-cone singularity is formed in both cases, but the  diffraction
is large for the neutrino and  small  for the charged lepton.

The probability of detecting the muon depends on the boundary condition
of the neutrino. When the neutrino  is detected at $\text{T}_\nu$, the muon
spectrum includes the diffraction component, but when the neutrino is not
detected,  the muon spectrum does not include the component. The latter
condition is 
standard, and the former is non-standard but may be verified
experimentally. 

 In   case of three masses $m_{{\nu}_i}$, and a mixing matrix
 $U_{i,\alpha}$, the diffraction term for  an $\alpha$  flavor neutrino  
is expressed 
as $\sum_i \tilde g(\omega_{\nu_i}\text{T})|U_{i,\alpha}|^2$, whereas
 the normal term is expressed  as  $|\sum_i U_{i,\mu}D(i)U^{\dagger}_{i,\alpha}|^2$, where  $i$ is  the mass
eigenstate,  $\alpha$ is the flavor eigenstate, and $D(i)$ is the free
wave of $m_{\nu_i}$. Hence, the diffraction term depends on the average
 mass-squared, $\bar m_{\nu}^2$,
but the normal term depends on  mass-squared differences, $\delta
 m^2_\nu$. At $\text{L} \rightarrow \infty$, the
diffraction term disappears, and  the normal terms  become  constants in the
 mass parameter region of the current study, $ \bar m_{\nu}^2 \gg \delta
 m_{\nu}^2 $.

Neutrino diffraction is different from  the diffraction of light
passing  through a hole. 
For the neutrino, the diffraction
pattern is formed   in a  direction parallel to the momentum  with the 
 phase difference $\omega_{\nu} \delta t $ of the non-stationary wave. The 
size of the pattern is determined by $\omega_{\nu}$, which  is
extremely small and stable with  variations in parameters. 
For light,  the diffraction pattern is formed  in a direction perpendicular
to the momentum with the phase difference $\omega_\gamma^{dB} \delta t $ of the 
stationary wave, where $\omega_\gamma^{dB}=c|\vec{p}_\gamma|/\hbar $.
The   shape of the pattern  is determined by  $\omega_\gamma^{dB}$, which is 
large and varies rapidly with  light's  energy. Thus, for observation,
   fine-tuning of  initial energy is necessary  in case of 
light but unnecessary in case of  neutrino.

{\bf 5 Summary and implications.}

 We presented a new mechanism for diffraction 
due to a many-body interaction in the decay region where the 
parent and daughters  overlap. The  probability to  detect the neutrino   
is given in Eq.\,$(\ref{probability-3})$, 
where  $G_0$ is  the normal term,   and $\tilde g(\omega_{\nu} \text T)$  
slowly 
decreases   with $\text{T}$.  The former agrees
with the  standard value obtained by an S-matrix of plane waves,
whereas 
the latter is a new 
term that  can be computed  by  $S[\text{T}]$ and
has its origin in diffraction due to   waves at finite t. In the
many-body state consisting of the pion, neutrino, and muon, the overlap
of the wave functions gives    a finite-interaction  
energy in $t \leq \tau$.  Because  kinetic energy is 
the difference between  total  and  interaction energies, 
it   varies with time. Consequently,  this  
many-body state  becomes non-uniform in space and time and shows 
a diffraction  pattern  that 
is  unique in the non-asymptotic region. 
This 
diffraction pattern is determined by the   
difference of the angular velocities,
$\omega_{\nu}=\omega_\nu^E-\omega_\nu^{dB}$, 
where $\omega_\nu^E={E_\nu/\hbar}$ and
$\omega_\nu^{dB}={c|\vec{p}_\nu|/\hbar}$. The term $\omega_{\nu} $
becomes  
an extremely small  value equal to $  {m_{\nu}^2c^4}/{(2E_{\nu} \hbar) }$  for 
neutrinos because of the  
unique  features  of neutrinos\cite {particle-data,Tritium,WMAP-neutrino}.
Consequently, the  diffraction term  becomes finite in a macroscopic 
spatial region $r \leq {2\pi E_{\nu} \hbar c}/{(
m_{\nu}^2c^4)}$ and affects experiments in a mass-dependent manner  at 
near-detector regions.
The area of this region  is exceptionally  large for  neutrinos. Waves
 accumulating  at the velocity of light   form  the light-cone
 singularity, which is 
 peculiar in relativistic
invariant systems, and exhibit   neutrino diffraction.

Neutrino diffraction   gives 
 new corrections to neutrino fluxes but not to the fluxes  of charged
 leptons;
 thus, it   is consistent with  all previous experiments involving
 charged leptons. 
The new term  has  various implications for existing neutrino anomalies and
future experiments. 
One anomaly is an excess of  neutrino flux at the near-detectors of ground 
experiments. Fluxes measured by the near detectors  of K2K
\cite{excess-near-detectorK2K} and MiniBooNE
\cite{excess-near-detectorMini}   show  excesses of 
$10\% $-$20\%$  in  Monte Carlo estimations,  
whereas the excess is not clear in MINOS 
\cite{excess-near-detectorMino}. These excesses  may be connected with the
diffraction component. With additional 
statistics, quantitative analysis  might become  possible   to test the
diffraction term.   
 Another anomaly is  LSND   \cite {LSND} in which   electron 
neutrinos in  pion decays have excesses.
Because  diffraction occurs  in the non-asymptotic region, 
 helicity suppression does not work. An electron mode is studied with
 a $(V-A)\times(V-A)$ current interaction in
 \cite{LSND-ishikawa-tobita}, and  it is found that   excess  in 
 near-detector regions is attributed to the diffraction component. 
The controversy between  LSND and others  is resolved.  
 Finally,  a new method that involves consideration of   
the distance or energy dependence of  neutrino flux 
may be developed  for determining the absolute neutrino mass.

Thus  neutrino diffraction  is visible at   macroscopic 
distances and can be confirmed  with near-detectors. At much larger
distances than that mentioned above,  the diffraction component
disappears, and only 
the normal component, including  the neutrino flavor oscillation, 
remains. If  masses do not satisfy   ${\bar m_\nu}^2 \gg \delta
m_\nu^2$ but satisfy  ${\bar m_\nu}^2 \approx \delta m_\nu^2$, 
then the neutrino fluxe behaviors are more  complicated. 

A new quantum phenomenon of neutrinos on a  macroscopic 
scale  due to  the  many-body weak
interaction  was derived, and its  physical quantity 
determined by the 
absolute neutrino mass  was presented.
 
In this study, we used the Hamiltonian expressed by the pion field and 
 neglected  higher-order effects such as  the pion mean-free-path
and the unified gauge 
theory.   The interaction of $(V-A) \times (V-A)$ does not modify
 the result on the muon mode but modifies  the electron mode, and other
 higher-order effects do not give a correction. We will
study these problems  and other large-scale physical phenomena 
of low-energy neutrinos in subsequent studies.

{\bf Acknowledgements.}  This study  was partly supported by a 
Grant-in-Aid for Scientific Research (Grant No. 24340043).
The authors  thank Dr. Nishikawa
and Dr. Kobayashi for useful discussions on 
the near detector of the T2K experiment, and Dr. Asai, Dr. Kobayashi, Dr. Mori, and Dr. Yamada
for their useful inputs  on interferences. 
{}


\begin{thebibliography}{}

\bibitem{MSW-1} L.~Wolfenstein, Phys. Rev. \textbf{D17}, 2369(1978).
\bibitem{MSW-2} S.~P.~Mikheev and A.~Yu.~Smirnov, Sov.~J.~Nucl.~Phys. \textbf{42}, 913 (1985); Nuovo~Cim. \textbf{C9},~17(1986).
\bibitem{LSZ} H.~Lehman, K.~Symanzik, and W.~Zimmermann, Nuovo~Cimento.\textbf{1},~205(1955).

\bibitem{Low} F.~Low, Phys. Rev. \textbf{97}, 1392(1955).



\bibitem{Ishikawa-Shimomura} K. Ishikawa and T. Shimomura.
	Prog. Theor. Phys.  \textbf{ 114}, 1201(2005) [hep-ph/0508303].
\bibitem{Ishikawa-Tobita-ptp} K. Ishikawa and Y. Tobita.
	Prog. Theor. Phys.  \textbf{ 122}, 1111(2009) [arXiv:0906.3938[quant-ph]]. 

\bibitem{Ishikawa-Tobita} K. Ishikawa and Y. Tobita. 
AIP Conf. proc. \textbf{1016}, 329(2008);
arXiv:0801.3124 [hep-ph].

\bibitem{Kayser} B.~Kayser,~Phys.~Rev.~D\textbf{24},~110(1981);~Nucl.Phys.~B\textbf{19}~(Proc.Suppl),~177(1991).

\bibitem{Giunti} C.~Giunti, C.~W.~Kim, and U.~W.~Lee, Phys. Rev. \textbf{D44}, 3635(1991).
\bibitem{Nussinov} S.~Nussinov, Phys. Lett. \textbf{B63}, 201(1976).
\bibitem{Kiers} K.~Kiers, S.~Nussinov and N.~Weiss. 
	Phys. Rev. \textbf{D53}, 537(1996) [hep-ph/9506271].
\bibitem{Stodolsky} L.~Stodolsky.  Phys. Rev. \textbf{D58}, 036006(1998) [hep-ph/9802387].

\bibitem{Lipkin} H.~J.~Lipkin.~Phys.~Lett.~B\textbf{642},~366(2006) [hep-ph/0505141].

\bibitem{Akhmedov} E. K.  Akhmedov.~JHEP. \textbf{0709}, 116(2007) [arXiv:0706.1216 [hep-ph]].

\bibitem{Asahara} A.~Asahara, K.~Ishikawa, T.~Shimomura, and T.~Yabuki,
Prog. Theor. Phys. \textbf{113}, 385(2005) [hep-ph/0406141]; T.~Yabuki and K.~Ishikawa.
Prog. Theor. Phys. \textbf{108}, 347(2002).

\bibitem{particle-data}
  J. Beringer {\it et al.}  [Particle Data Group],
Phys. Rev. \textbf{D86}, 010001 (2012).
\bibitem{Tritium}
V.~N.~Aseev et al.~Phys.~Rev.~D\textbf{84},~112003(2011)
[arXiv:1108.5034[hep-ex]].

\bibitem{WMAP-neutrino}
E. Komatsu, et al.~Astrophys. J. Suppl.~\textbf{192},~18(2011) [arXiv:1001.4538[astro-ph.CO]].




\bibitem{Wilson-OPE} K.~Wilson, in Proceedings of the Fifth International Symposium on Electron and Photon Interactions at High Energies, Ithaca, New York, 1971,~115~(1971).
See also  N.~N.~Bogoliubov and D.~V.~Shirkov, 
{\it Introduction to the Theory of Quantized  Fields}~(John Wiley \&
	Sons, Inc. New York, 1976).
\bibitem{Ishikawa-tobita-lightcone}
 K.~Ishikawa and Y.~Tobita. arXiv:1206.2593,1209.5586[hep-ph].  
\bibitem{excess-near-detectorK2K}
M.~H.~Ahn, et~al. ~Phys. Rev. D\textbf{74},~072003(2006) [hep-ex/
0606032].

\bibitem{excess-near-detectorMini}
A.~A.~Aguilar-Arevalo, et~al,~Phys. Rev. D\textbf{79},~072002,~(2009).

\bibitem{excess-near-detectorMino}
P.~Adamson, et~al. ~Phys. Rev. D\textbf{77},~072002(2008) [arXiv:0711.0769[hep-ex]].

\bibitem{LSND}
C.~Athanassopoulos, et~al.~Phys. Rev. Lett. \textbf{75},~2650(1995) [nucl-ex/9504002];
  \textbf{77},~3082(1996) [nucl-ex/9605003];
	\textbf{81},~1774(1998) [nucl-ex/9709006].

\bibitem{LSND-ishikawa-tobita}
 K.~Ishikawa and Y.~Tobita. arXiv:1109.3105 [hep-ph].  
\end{thebibliography}
\end{document}